\documentclass[useAMS,usenatbib]{mn2e}

\title[Slow modes in rotating B-type stars]
{Excitation and visibility of slow modes in rotating B-type stars}
\author[W. A. Dziembowski, J. Daszy\'nska-Daszkiewicz, A. A. Pamyatnykh]
{W. A. Dziembowski$^{1,2}$\thanks{E-mail: wd@astrouw.edu.pl},
J. Daszy\'nska-Daszkiewicz$^{3}$ and A. A. Pamyatnykh$^{2,4}$\\
$^{1}$Warsaw University Observatory, Al. Ujazdowskie 4, 00-478
Warsaw, Poland\\ $^{2}$Copernicus Astronomical Center, Bartycka
18, 00-716 Warsaw, Poland\\ $^{3}$Instytut Astronomiczny,
Uniwersytet Wroc{\l}awski,
   ul. Kopernika 11, 51-622 Wroc{\l}aw, Poland\\
$^{4}$Institute of Astronomy, Russian Academy of Sciences,
   Pyatnitskaya Str. 48, 109017 Moscow, Russia}

\usepackage[centertags]{amsmath}
\usepackage{amsfonts}
\usepackage{amssymb}
\usepackage{graphicx}
\newcommand{\am}{\vert m \vert}
\newcommand{\ve}{\mbox{\boldmath{$e$}}}
\newcommand{\vnab}{\mbox{\boldmath{$\nabla$}}}
\newcommand{\vv}{\mbox{\boldmath{$v$}}}
\newcommand{\vn}{\mbox{\boldmath{$n$}}}
\newcommand{\vxi}{\mbox{\boldmath{$\xi$}}}
\newcommand{\vzeta}{\mbox{\boldmath{$\zeta$}}}

\begin{document}

\date{Accepted 2006 October 2; Received 2006 September 29; in original form 2006 July 7}

\pagerange{\pageref{firstpage}--\pageref{lastpage}} \pubyear{2006}

\maketitle

\label{firstpage}

\begin{abstract}
We use the traditional approximation to describe oscillations with frequencies comparable
to the angular rotation rate. Validity of
this approximation in application to  main-sequence B stars is discussed.
Numerical results regarding mode stability and
visibility are presented for a model of the Be star HD 163868.
For this object, Walker et al.\,(2005)
detected a record number of mode frequencies using
data from the small space telescope {\it MOST}.
Our interpretation of these data differs from that of Walker et
al. In particular, we interpret peaks in the lowest frequency range
as retrograde g modes. We find instability in a large number of modes
that remain undetectable because of unfavourable aspect and/or effect of
cancellation. There is no clear preference to excitation of
prograde modes.
\end{abstract}


\section{Introduction}

Results of linear stability calculations suggest that excitation
of high-order g modes in main-sequence B-type stars must be a
common phenomenon. The instability is typically robust. There is a
significant excess of driving over damping and many modes are
simultaneously unstable in individual stellar models. With the
standard Population I composition, the instability is found in
models of nearly all B-type stars. Yet, the observed pulsation
associated with the slow modes is unspectacular. Individual modes
never attain high amplitudes. Only in a small fraction of B stars
do we have a firm evidence for the presence of oscillations and
typically only few modes are seen. One possible reason why,
despite strong instability, we never see high amplitude pulsation
of this type is a collective saturation by a large number of modes
which are difficult to detect. Recent report by Walker et
al.\,(2005) on the detection of about sixty periods in the data on
HD 163868 obtained from {\it MOST} may support this idea.

The fact that HD 163868 is a rapidly rotating Be type object makes
the detection of many oscillation modes particularly interesting.
The cause of activity observed in stars of this type is still
not understood. Seismic constraints derived from mode frequencies
on the internal structure would be of great value in this context. Furthermore,
it is possible that oscillations play an active role in these
objects by inducing angular momentum transport, as first
suggested by Osaki (1986) and Penrod (1986) and more recently, in
the context of opacity-driven modes, by Lee (2006).
Let us add that an asymmetry in excitation of prograde and retrograde modes may
induce a net helicity of motion and thus an action of magnetic dynamo.

To extract seismic information, as well as to assess possible active role
of oscillation, we have to identify modes excited in the star.
The task of individual mode identification in an object like HD
163868 is formidable and will be attempted in this work. Like
Walker et al.\,(2005), we will only try to determine angular degrees and
azimuthal orders of the three groups of modes which are centered on
periods of 8 days, and 14 and 7 hours. These authors, however, have not
found unstable modes that could explain the longest period group.

We will argue that the observed modes represent only a small
subset of the excited modes. Other modes having similar rms
amplitudes may be not detectable because of cancellation and/or
unfavourable aspect. Our analysis of stability and visibility is
based on linear non-adiabatic calculations with the effects of
rotation  treated in the traditional approximation. The same
approximations were adopted by Townsend (2005a,b) in his
extensive survey of stability of slow modes  in models of rotating
main-sequence B-stars. The slow modes include high-order g modes
with frequencies comparable to the angular rate of rotation and
certain r modes (called mixed gravity-Rossby modes), which at
sufficiently high rotation rate become propagatory in the
radiative envelopes. Both types are candidates for
identification of modes excited in HD 163868.

Our identification differs from that of Walker et al.\,(2005). One
may suspect that the difference should be blamed to our use of the
traditional approximation, which they avoid. However, as we
explain in the next section, the approximation should be valid in
the application to, at least, certain modes in HD 163868.

In Section 3, we present results on stability of slow modes in a
model of HD 163868 which is very similar to the model used by
Walker et al. (2005). We review properties of unstable modes in this and
similar models and compare our results with those of our
predecessors. In Section 4 we quantify the dependence of
observable amplitude on the mode geometry and the aspect angle for
the most unstable low-degree modes. Comparison with observations
is presented in Section 5.

\section{On the traditional approximation}

There is a number of recent papers where {\it the traditional
approximation} is used to describe properties of low frequency
modes in rotating stars (e.g. Bildstein, Ushomirsky \& Cutler
1996; Lee \& Saio 1997;  Townsend 2003a; Savonije 2005; Townsend
2005a,b). Readers are referred, in particular, to Townsend (2003a)
for details of derivation of simplified equations for adiabatic
oscillations known as {\it Laplace's Tidal Equations}, which
employ the traditional approximation. The last three papers extend
the approximation to the case of non-adiabatic oscillation. Here,
we focus on the validity of approximation and its extension to the
case of non-uniform rotation and we will quote only the most
essential equations.

The generally adopted, initial simplifications of the linearized
Euler equations,
\begin{equation}
\left({d\vv\over dt}\right)'+{1\over\varrho}\vnab p'-
{\varrho'\over\varrho^2}\vnab p+\vnab \Phi'=0,
\end{equation}
where all notation is standard, are
(i) slow and (ii) uniform rotation, (iii) oscillation frequency, $\omega$, of the same order
as the angular velocity of rotation, $\Omega$, (iv) neglect of $\Phi'$
(the Cowling approximation), (v) adiabaticity of perturbations.
Here we adopt (i) and (iii), that is, we assume
\begin{equation}
{\omega^2r\over g}\sim{\Omega^2r\over g}\equiv\epsilon\ll1,
\end{equation}
but not (ii) as we allow the $r$ dependence in $\Omega$.
Indeed, the traditional approximation applies to this more general case
in the essentially unchanged form.
We begin assuming (v) but later we will consider non-adiabatic
effects, which are in fact of main interest for us in this paper.

Inequality (2) implies that the effects of centrifugal distortion
may be neglected which together with (iv) reduces Eq.\,(1) to
\begin{equation}
\left({d\vv\over dt}\right)'+{1\over\varrho}\vnab p'+
{\varrho'\over\varrho}g\ve_r=0.
\end{equation}
Since we allow non-uniform rotation, the constant eigenfrequency,
$\omega_{\rm obs}$, must be calculated in the inertial system, while
$\omega$ - the frequency calculated in the local corotating system
- may vary with $r$.
The  acceleration in the inertial system expressed in terms of the displacement
vector,
\begin{equation}
\vxi\equiv r\vzeta(r,\vartheta)\exp[{\rm i}(m\varphi-\omega_{\rm obs} t)],
\end{equation}
is given by
\begin{equation}
\left({d\vv\over dt}\right)' =-\omega^2\vxi-2{\rm
i}\Omega\omega\ve_z\times\vxi+
(\ve_r\sin\vartheta+\ve_\vartheta\cos\vartheta) {d\Omega^2\over
dr}r\sin\vartheta\xi_r,
\end{equation}
where $\omega(r)=\omega_{\rm obs}-m\Omega(r)$. Note that with the
adopted form of the time dependence, the modes of azimuthal order
$m>0$ are {\it prograde} and those with $m<0$ are {\it
retrograde}. This is opposite to what has been adopted in the
cited papers employing traditional approximation  but it is in
agreement with the {\it {\c C}e{\c s}me Resolution} (see Appendix
to Christensen-Dalsagaard \& Dziembowski 2000).

Using the adiabaticity condition
\begin{equation}
{\varrho'\over\varrho}={1\over\Gamma_1}{p'\over p}+A{\xi_r\over r},
\end{equation}
where we denoted
$$A=-{d\ln\varrho\over d\ln r}-V_g\quad\mbox{ and } V_g={g\varrho r\over p\Gamma_1},$$
and introducing one additional eigenfunction, $\zeta_p(r,\vartheta)$, through the
following identity
\begin{equation}
p'=gr\varrho \zeta_p(r,\vartheta)\exp[{\rm i}(m\varphi-\omega_{\rm obs}
t)],
\end{equation}
we get from Eq.\,(3)

\begin{eqnarray}
\left(N^2-\omega^2+{d\Omega^2\over
dr}r\sin^2\vartheta\right)\zeta_r + 2{\rm
i}\Omega\omega\sin\vartheta \zeta_\varphi&+&~\nonumber\\
{g\over r}\left(r{\partial\over\partial r}+U-A-1\right)\zeta_p
&=&0,
\end{eqnarray}

\begin{equation}
{d\Omega^2\over dr}r\sin\vartheta\cos\vartheta
\zeta_r-\omega^2\zeta_\vartheta+2{\rm i}\Omega\omega\cos\vartheta \zeta_\varphi+
{g\over r}{\partial \zeta_p\over\partial\vartheta}=0,
\end{equation}
and
\begin{equation}
2\Omega\omega\sin\vartheta \zeta_r+2\Omega\omega\cos\vartheta
\zeta_\vartheta- {\rm i}\omega^2\zeta_\varphi-{mg\over
r\sin\vartheta}\zeta_p=0.
\end{equation}
In Eq.\,(8) we use standard notation $N$ for the Brunt-V\"ais\"al\"a
frequency and $U$ for the logarithmic derivative of the fractional
mass.
The continuity equation combined with Eq.\,(6) yields
\begin{equation}
\left(r{\partial\over \partial r}+3-V_g\right)\zeta_r+
\left({\partial\over\partial\vartheta}+\cot\vartheta\right)\zeta_\vartheta+
{{\rm i}m\over\sin\vartheta}\zeta_\varphi+V_g\zeta_p=0,
\end{equation}
which closes the system of equations for the eigenvalues $\omega_{\rm obs}$ and the
associated eigenfunctions $\zeta_r$, $\zeta_\vartheta$, $\zeta_\varphi$,
and $\zeta_p$.

In the traditional approximation all terms
arising from the Coriolis force, except those in Eqs.\,(9) and (10) which contain horizontal components
of $\vzeta$, are dropped out.
A simple justification, based on the local wave approximation,  was given by Lee and
Saio (1997). The approximation is valid if
\begin{equation}
N=\sqrt{gA\over r}\gg\Omega.
\end{equation}
Earlier, the same condition was
obtained by Dziembowski \& Kosovichev (1987), who did not rely on the
wave approximation and did not assume uniform rotation.
Conditions  (2) and (12) justify the neglect of those Coriolis terms
and the term containing the derivative of $\Omega$,
both in the outer evanescent zone (assuming that it is radiative)
as well as in the g-wave propagation zone. In both zones, we have
$|\zeta_\vartheta|\sim |\zeta_\varphi|\gg\max(|\zeta_r|,|\zeta_p|)$, which
justifies simplification of Eqs.\,(9) and (10).
The relative scaling of $\zeta_r$ and $\zeta_p$ is different in the
evanescent and g-wave zone. In the former, where $V_g$ and $A$ are
large, $|\zeta_r|\sim|\zeta_p|\sim\epsilon \zeta_\vartheta\sim\epsilon \zeta_\phi$
and the simplification in Eq.\,(8) follows from $A\gg1$. In the
propagation zone, where the mode amplitude varies rapidly, we have
$$\left|r{\partial \zeta_p\over\partial r}\right|\sim\sqrt{N\over\Omega}|\zeta_p|
\sim {A\over\epsilon}|\zeta_r|\gg|\zeta_\varphi|\gg |\zeta_r|,$$ which justifies the
neglect of all terms containing $\Omega$.
Note that the $\omega^2$ term in the coefficient at $\zeta_r$ could have
been ignored too but traditionally this is not done.

With the traditional approximation, the $r$ and $\vartheta$
dependence in the eigenfunctions may be separated. The latter
dependence is given by the Hough functions. Following Townsend
(2003a), we substitute in  Eqs.\,(8-11)
\begin{equation}
\zeta_r=y_1(r)\Theta(\vartheta),\qquad \zeta_p=y_2(r)\Theta(\vartheta)
\end{equation}
\begin{equation}
\sin\vartheta \zeta_\vartheta=z(r)\hat\Theta(\vartheta),
\quad \mbox{ and }\quad {\rm i}\sin\vartheta
\zeta_\varphi=z(r)\tilde\Theta(\vartheta)
\end{equation}
to obtain ordinary differential equations for $y_1$ and
$y_2$,

\begin{equation}
r{d y_1\over dr}=(V_g-3)y_1+\left({\lambda gr\over\omega^2}-V_g\right)y_2,
\end{equation}
and
\begin{equation}
r{d y_2\over dr}=\left({r\omega^2\over g}-A\right)y_1+(A+1-U)y_2,
\end{equation}
where,  for a uniform rotation, $\lambda$ is the separation parameter. The radial dependence
for the horizontal components of $\vxi$ is given by
\begin{equation}
z={g\over r\omega^2}y_2.
\end{equation}
Eqs.\,(15) and (16) look just like in the case of no rotation,
except that now we have $\omega=\omega_{\rm obs}-m\Omega$ and
in the place of $\ell(\ell+1)$ the separation parameter
$\lambda$, which is determined as an eigenvalue in the equations
for the Hough functions. We write these equations in the form
\begin{equation}
\tilde\Theta=-m\Theta+s\mu\hat\Theta
\end{equation}
\begin{equation}
(1-\mu^2){d\Theta\over d\mu}=-ms\mu\Theta+(s^2\mu^2-1)\hat\Theta
\end{equation}
\begin{equation}
(1-\mu^2){d\hat\Theta\over d\mu}=[\lambda(1-\mu^2)-m^2]\Theta+ms\mu\hat\Theta,
\end{equation}
where $\mu\equiv\cos\vartheta$ and $s\equiv2\Omega/\omega$ is
called {\it the spin parameter}. Eqs.\,(19-20) differ from
Townsend's Eqs.\,(21-22) only in the sign of $s$, which he denoted
$\nu$, in consequence of different sign convention for the time
dependence. The equations together with boundary conditions at
$\mu=0$ (symmetry) and $\mu=1$ (regularity) define the eigenvalue
problem on $\lambda$. The dependence of $\lambda$ on $s$ and
related properties of the Hough functions have been discussed in
great detail by, for example, Bildstein et al.(1996), Lee \& Saio (1997),
and by Townsend (2003a).

Of our interest here are only modes which are propagating in
radiative zones, where $A>0$, that is, corresponding to
$\lambda>0$. These are the g modes for which
$\lambda\rightarrow\ell(\ell+1)$ at $s\rightarrow0$, as well as
the mixed gravity-Rossby modes, for which $\lambda$ changes sign
from minus to plus at $s=|m|+1$ (for brevity, in this paper, we
shall call them r modes). It was shown, independently by Townsend
(2005b) and Savonije (2005) and confirmed by Lee (2006), who did
not use the traditional approximation, that such modes may be
driven in  B-type and early A-type stars.  Let us note that if
$\Omega$ is a function of $r$, then such modes may be trapped in
the region of rapid rotation where $\lambda>0$ while it is $<0$ in
the rest of the star interior.

In the case of massive main-sequence stars, the traditional approximation cannot be
used in their convective cores. However, low frequency waves do not propagate there and thus
the core surface may be treated as a boundary. Except
for zero-age main-sequence stars there is a nearly discontinuous  transition at the
core edge, $r=r_c$, from $N\gg\Omega$ to $N\approx0$. In the
core, $\zeta_p$ and other eigenfunctions are slowly varying with distance
from the center. Thus, with $N\approx0$, Eq.\,(8) implies
\begin{equation}
|\zeta_p|\sim\epsilon|\zeta_r|\quad\mbox{ for } r\le r_c.
\end{equation}
On the other hand, in the propagation zone, we
have $k_r\zeta_p\approx N^2\zeta_r/g,$
where the local radial wave number is given by
$k_r=k_HN/\omega$ and $k_H=\sqrt{\lambda}/r$ is the horizontal
wave number. Hence,
\begin{equation}
|\zeta_p|\sim \sqrt{\epsilon{A\over\lambda}}\quad\mbox{ for } r\ge r_c.
\end{equation}
Since there is no jump of density at $r=r_c$, the continuity of
$\delta p$ and $\xi_r$ imply the continuity of $\zeta_p$. Thus,
from Eqs.\,(21) and (22), we get the boundary condition consistent with the traditional approximation
to be imposed on solutions of Eqs.\,(15) and (16),
\begin{equation}
{y_2(r_c)\over y_1(r_c)}\sim\sqrt{\epsilon}\approx0.
\end{equation}
In stars of our interest, there are thin convective layers,
connected with the HeII ionization and, if $M\gtrsim7M_\odot$,
with the Fe opacity bump. Because of their small vertical extent, such
layers cannot affect significantly global mode geometry.

The two boundary conditions, which should be applied on solution of Eqs.
(15-16), are
\begin{equation}
y_2=0 \mbox{ at }r=r_c\quad\mbox{ and }\quad y_2=y_1\mbox{ at }r=R.
\end{equation}
The same boundary conditions remain valid when  non-adiabatic
effects are taken into account.

Eqs.(15-20) remain valid in the case of shellular rotation
but then, both $\omega$ and $\lambda$ must be regarded as
functions of $r$. The derivation given by Dziembowski \&
Kosovichev (1987) employed expansion of the $\theta$ dependence in
a series of the associated Legendre functions. Assuming
$\Omega/N\sim\delta\ll1$, they decomposed the infinite system of
differential equations in $r$ into asymptotically independent
second-order equations equivalent, in the limit
$\delta\rightarrow0$, to Eqs.(15-16). In their derivation there is
an inconsequential omission of terms $\sim\delta d\ln\Omega/d\ln
r$ and an implicit assumption that the logarithmic derivative of
$\Omega$ is of the order of 1.

Townsend (2005a) and Savonije (2005) extended the traditional
approximation to the case of non-adiabatic oscillations. Using the
same equation as in the case of no rotation with only $\lambda$ in
place of $\ell(\ell+1)$ is incorrect in the expression for the
horizontal heat losses. However, these losses, which are
proportional to $k_H^2$, make rather small contribution to the
overall work integral. Another simplification is the use of
$\lambda$ with $\omega$ calculated in the adiabatic approximation.
This is well justified in our applications  because the
non-adiabatic change in $\omega$ is very small.

Lee \& Saio (1989) pointed out that even at $\delta\ll1$
coupling between modes of close frequencies, the same $m$ and
symmetry may have a significant effect. In a specific example of
$4M_\odot$ star sequence, Lee (2001) found that some modes, which
are unstable if the traditional approximation is used, are found
stable if the approximation is abandoned and the truncated
expansion in Legendre functions is used for the eigenfunctions.
However, many modes remained unstable and we expect that this
applies also in the case of models considered in this paper. We
will discuss this matter further in Section 3.
\begin{figure*}
\centering
\includegraphics[width=140mm, clip]{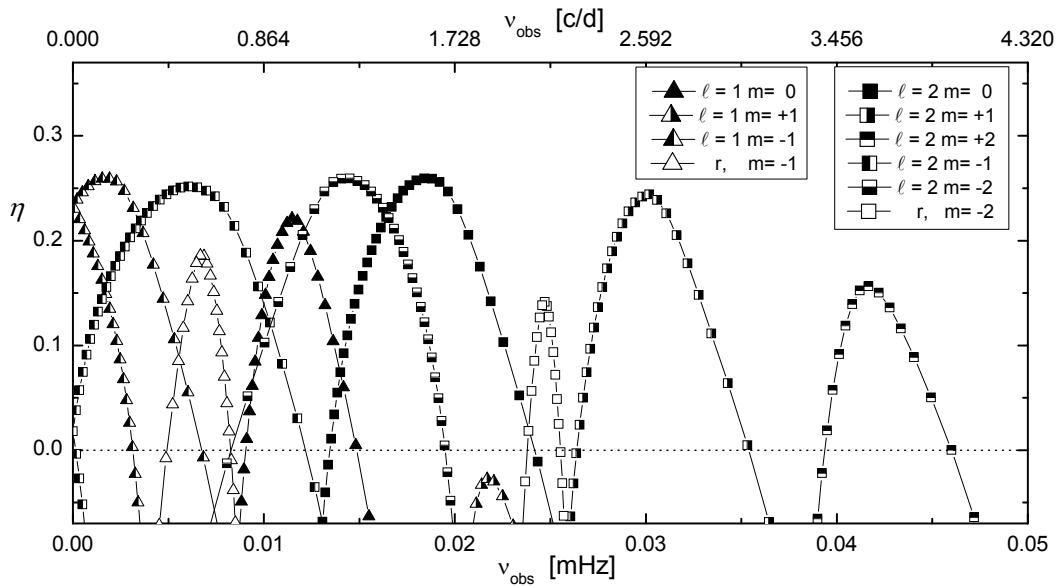}
\caption{The instability parameter, $\eta$ (see Eq.(25)) as function of frequency
for slow modes of low azimuthal orders in the model of HD 163868. The frequencies are calculated in
the inertial system. The frequency of rotation is $\nu_{\rm rot}\equiv\Omega/2\pi=1.42$ c/d.}
\end{figure*}

\section{Unstable modes}

With the traditional approximation, stability properties of slow
modes are determined by two parameters: $\lambda$ and $\omega$. In
this section, we assume uniform rotation, hence $\omega$ is the
mode eigenfrequency in the corotating system. The whole influence
of rotation and the azimuthal order is absorbed in the
$\lambda(s)$ dependence. Thus, like in the case of no rotation, we
deal with a two-dimensional instability range for a specified
model. In the present case, the two parameters are $\omega$ and
$\sqrt{\lambda}/\omega$. The latter determines radial order of the
mode, $n$, and the $r$ dependence of the Lagrangian pressure
perturbation, $\delta p/p$. For instability, the amplitude of the
perturbation should be large and slowly varying in the layer of
the iron opacity bump. The frequency, $\omega$, determines
resulting perturbation of the radiative flux. Furthermore, a match
is needed between the thermal time-scale of this layer, $\tau_{\rm th}$,
and the pulsation period, $\Pi=2\pi/\omega$.
\begin{table*}
\centering
\caption{Parameters of the most unstable modes with $|m|\le2$ for
the stellar model with $M=6.0 M_{\odot}$, $\log T_{\rm eff}= 4.226$, $\log L/L_{\odot}=3.094$,
and a rotational velocity of 300 km/s, which corresponds to $\nu_{rot}=1.4165$ c/d.
See Eq.(29) for the definition of $f$.  }
\begin{tabular}{|ccccccccrrc|}
\hline
$\ell$ & $m$ & $n$ & type & parity & $\nu_{\rm obs}$ [c/d] & $\nu_{\rm star}$ [c/d] & spin & $\lambda~~$ & $\eta~~~$ & $f$ \\
\hline
  1 &  0 & 20 & zonal, g & odd  & 0.9911 & 0.9911 & 2.86 &  8.76 &  0.221 &  (7.67, 12.96) \\
\hline
  1 &  1 & 15 & progr, g & even & 1.8748 & 0.4582 & 6.18 &  1.09 & -0.027 &  (9.46,  9.00) \\
\hline
  1 & -1 & 21 & retro, g & even & 0.1654 & 1.5819 & 1.79 & 24.69 &  0.259 &  (1.15,  8.93) \\
\hline
    & -1 & 19 & retro, r & odd  & 0.5772 & 0.8393 & 3.38 &  5.70 &  0.186 &  (8.94, 12.68) \\
\hline
\hline
  2 &  0 & 20 & zonal, g & even & 1.5893 & 1.5893 & 1.78 & 27.29 &  0.259 &  (1.78,  9.70) \\
\hline
  2 &  1 & 20 & progr, g & odd  & 2.6081 & 1.1916 & 2.38 & 12.71 &  0.244 &  (4.65, 11.58) \\
\hline
  2 &  2 & 18 & progr, g & even & 3.5997 & 0.7666 & 3.70 &  4.28 &  0.156 &  (8.69, 11.88) \\
\hline
  2 & -1 & 23 & retro, g & odd  & 0.5228 & 1.9393 & 1.46 & 45.07 &  0.252 &  (-0.51, 6.60) \\
\hline
  2 & -2 & 22 & retro, g & even & 1.2515 & 1.5816 & 1.79 & 27.02 &  0.259 &  (1.85,  9.78) \\
\hline
    & -2 & 19 & retro, r & odd  & 2.1348 & 0.6983 & 4.06 &  3.94 &  0.142 & (11.71, 12.93) \\
\hline
\end{tabular}
\end{table*}

The instability ranges in terms of $\ell$ and $\omega$ for
non-rotating main-sequence B-stars were discussed in detail by
Dziembowski, Moskalik \& Pamyatnykh (1993) and by Pamyatnykh
(1999). A corresponding discussion for rotating stars was
presented recently by Townsend (2005a). For a family of stellar
models, $\sqrt{\lambda}/\omega$ does not determine mode order but
still, to a large extent, determines the shape of the pressure
eigenfunction in the outer layers, which bring the main
contribution to the work integral. Thus, the evolutionary radius
increase results in longer periods of unstable modes. The
associated decrease of the effective temperature, causing an
increase of $\tau_{\rm th}$, acts in the same direction. This is
why the initial effect of evolution is an enhancement of
instability. The tendency is reversed only near the end of the
main-sequence evolution when dissipation in the g-wave propagation
zone becomes significant. Too small values of $\tau_{\rm th}$ are
the cause why the modes with low values of $\lambda$ are stable in
massive ($M\gtrsim4.5M_\odot$) hot objects.

In this paper we present numerical results for a single stellar
model calculated with Warsaw-New Jersey stellar evolutionary code  (see
e.g. Pamyatnykh 1999) assuming no mixing beyond the convective
core and including only mean effect of the centrifugal force. The same
approximations were adopted by Walker et al.\,(2005). Our model is
characterized by the following parameters:
 $M=6M_\odot$, $X_0=0.7$, $Z=0.02$,  $X_c=0.307$,
 $\log T_{\rm eff}=4.226$, $v_{\rm rot}=$300 km/s,
which are very similar to those used by these authors.

Our non-adiabatic pulsation code uses the same
approximation as adopted by Townsend (2005a) except that we apply
the inner boundary condition $y_2=0$ at the edge of convective
core. The code is almost a trivial modification of our standard
nonradial pulsation code in its version
adopting the Cowling approximation, which
is indeed very accurate for all considered modes.
In the adiabatic part of the code values of $\lambda$ are interpolated
from the tables of $\lambda(s)$ prepared with a separate code for
all types of modes of our interest. The values were not changed
in the non-adiabatic part.
We focused the main attention on modes which may be driven by the
opacity mechanism and are potentially detectable by photometry.
Thus, the modes considered have $\lambda$ greater than zero but not too large.
For each $\am\le 2$, we included g modes with the lowest
values $\lambda$. Sequences of modes of different radial order are denoted with $(\ell,m)$,
where $\ell(\ell+1)=\lambda(0)$. We also included
two sequences of lowest order retrograde r modes with $\lambda>0$ for $s>\am+1$.
The mode instability is characterized by the normalized work integral
\begin{equation}
\eta={W\over\int_0^R|{dW\over dr}|dr},
\end{equation}
where $W$ is the usual work integral.
The growth rate, $\Im(\omega)$, has the same sign
but its value is affected mostly by mode inertia. The
value of $\eta$, which varies between -1 and 1, is a better measure of
robustness of the instability and a better predictor of the mode
amplitude.
Fig.\,1 shows the values of $\eta$ for all considered modes in
their range of instability and close to it.
Table 1 provides more information about most unstable modes
(largest $\eta$).

The instability range in the observed frequencies, $\nu_{\rm obs}=\omega_{\rm obs}/2\pi$,
extends from nearly zero for retrograde $\ell=1$ g modes to 4 c/d for prograde $\ell=2$
modes. The former modes have frequencies in the vicinity of $\nu_{\rm
rot}=\Omega/2\pi$. The modes with frequencies $\nu_{\rm star}=\omega/2\pi<\nu_{\rm rot}$ are seen as
prograde modes. The range of frequencies in the corotating system is much narrower. We may
see in Table 1 that the most unstable modes occur near $\nu_{\rm star}\approx1.6$ c/d,
$\lambda\approx30$ and $n=20$. The prograde
$\ell=1$ modes have $\lambda<2$ and, thus, at $\sqrt{\lambda}/\omega$ values favourable for
driving, too long periods. Such modes were found unstable in
somewhat cooler models. A lower value of $T_{\rm eff}$ would also result
in lower $\nu_{\rm star}$ of the unstable modes. The shift in
$\nu_{\rm obs}$ is the same but, if $m\nu_{\rm rot}>\nu_{\rm
star}$, of opposite sign. In the vicinity of our model a shift of
0.01 in $T_{\rm eff}$ leads to the 0.1 c/d frequency shift of most unstable
modes. A change in the adopted value of $v_{\rm rot}$ would
shift the instability ranges in $\nu_{\rm obs}$ for $m\ne 0$ modes.

There are many more strongly unstable modes than shown in Fig.\,1.
The instability continues up to $\lambda\approx600$. Modes with
$\eta>0.2$ are found up to $\lambda\approx120$, which corresponds
to $m=11$ for sectorial prograde modes. Owing to higher $\lambda$
values such modes are subject to much larger observable amplitude
reduction. Visibility of modes is the subject of the next section.
Here, we just want to point out that we should expect in any of
multiperiodic B-type star many more excited modes than those
detected.  Such modes have properties similar to those listed in
Table 1, except they would have much lower surface averaged
amplitudes. Possible excitation of many unobservable modes must be
kept in mind while considering the feedback of oscillations on
stellar rotation. The results shown in Fig.\,1  and in Table 1
might suggest that there is preference to retrograde mode
excitation but it is not found in the full set of the unstable
modes. For instance, $\eta\ge0.2$ was found in over 450 prograde
modes, and in less than 370 retrograde modes. Certainly, there is
no evident asymmetry in driving prograde and retrograde modes.

Our findings regarding mode instability differ from that presented
by Walker et al.\,(2005). Unlike these authors, we do not find
instability of the prograde $\ell=1$ modes. On the other hand, we
find many retrograde $\ell=1$ modes, which may account for the
longest periods measured in HD 163868, whereas no instability of
such modes was found in that work. The general pattern of the
difference is that we find instability in many more modes with
higher $\lambda$. Perhaps the traditional approximation, which we
rely on, is not applicable. The mode coupling, as described
by Lee (2001), may stabilize only certain modes. Could it
stabilize all retrograde $\ell=1$ modes and destabilize all
prograde $\ell=1$ modes? We do not know the answers. Perhaps,
there is a difference in internal structure between the models
used by us and that used by Walker et al.\,(2005). We should
also be aware that the Lee \& Saio method, which is based on a
truncated series expansion, used in that paper, may not be
accurate.

Considering possible identification of modes detected in  HD
163868, we should take into account visibility conditions for
modes described by the Hough functions.

\section{Visibility}
Observed amplitude of a mode depends on its true amplitude,
geometry and the aspect angle, $i$. Photometric amplitudes for
modes described by single spherical harmonics were calculated in
a number of papers. Results of corresponding calculations for modes described
by the Hough functions were first published by Townsend (2003b), who
relied on truncated expansion of $\Theta$ in associated Legendre
functions. In our calculations of the photometric and radial
velocity amplitudes, we avoid such expansion and carry out
two-dimensional integration over visible hemisphere.
Details of our approach and a survey of the results will be
published elsewhere (Daszy\'nska-Daszkiewicz et al., in preparation). Here we
quote only the background formulae and present discussion of
visibility of modes listed in Table 1 as a function of aspect.
\begin{figure*}
\centering
\includegraphics[width=140mm, clip]{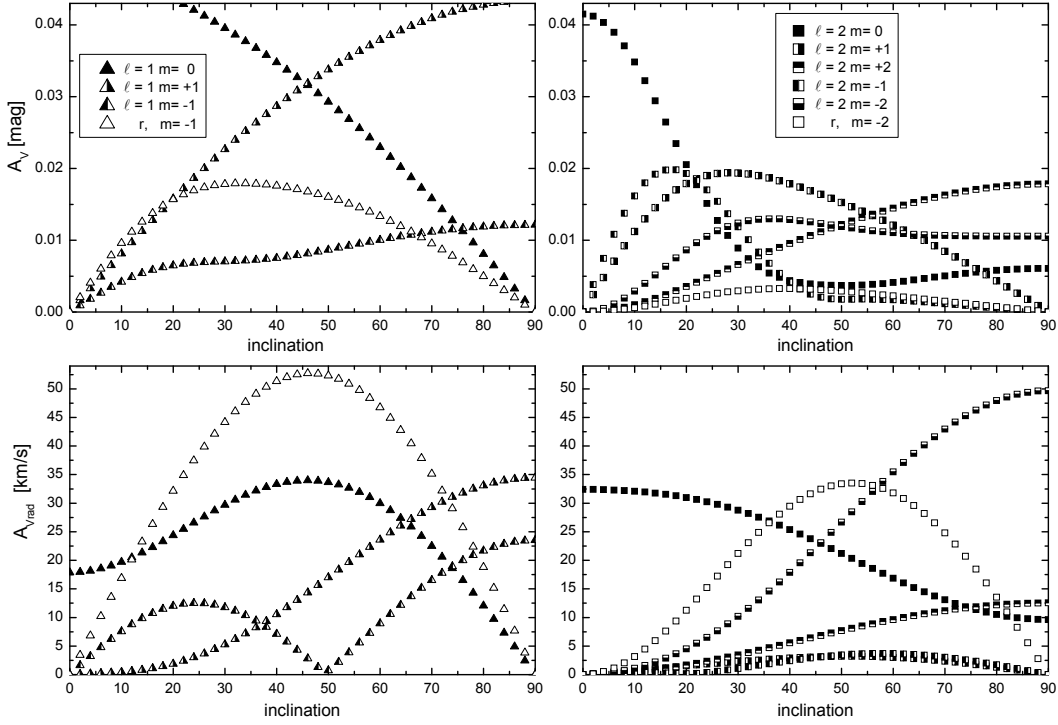}
\caption{Light amplitudes in the Geneva $V$ band (upper panels)
and in radial velocity (lower panel) as functions of the
inclination angle for modes listed in Table 1.}
\end{figure*}

Within the linear non-adiabatic theory,
there is a freedom in normalization of eigenfunctions. Here we
adopt the most common one, that is $y_1(R)=1$, and
$$\int_0^1\Theta^2d\mu=1.$$
Thus, we write
\begin{equation}
\xi_r(R)=\varepsilon R\Theta Z,
\end{equation}
where $Z={\rm exp}[{\rm i}(m\varphi-\omega t)]$.
The value of $|\varepsilon|$ (not to be confused with $\epsilon$ in Section 2)
determines true mode amplitude.
Note that the surface displacement is given by $\Re[\xi_r(R)]$ and its rms
value is equal to  $|\varepsilon|/\sqrt{4\pi}$.
The surface boundary condition,
$y_2(R)=y_1(R)$, combined with Eqs.\,(4) and (14)
leads to the following expressions for the horizontal components of the
displacement
\begin{equation}
\xi_{\theta}=\varepsilon\frac{GM}{\omega^2R^2}\frac{\hat\Theta}{\sin\vartheta}Z
\end{equation}
and
\begin{equation}
\xi_{\varphi}=-{\rm i}\varepsilon\frac{GM}{\omega^2R^2}
\frac{\tilde\Theta}{\sin\vartheta}Z.
\end{equation}
These components are needed only for calculation of
the pulsation velocity field.

For evaluation of the photometric amplitudes, we need a
linear relation between the radial components of displacement and
perturbed flux which we write in the form
\begin{equation}
\frac{\delta {\cal F}_{\rm bol}}{{\cal F}_{\rm bol}}
\equiv4\frac{\delta T_{\rm eff}}{T_{\rm eff}}
= \varepsilon f\Theta Z,
\end{equation}
where $f$ is a complex coefficient determined from linear
non-adiabatic calculations. Its values for modes considered are given in the last column
of Table 1.
In order to calculate the flux in
a specified band, $x$,
we need also an  expression for the perturbed surface gravity,
which is
$$\frac{\delta g}{g} = -\varepsilon \left({r\omega^2\over
g}+2\right)\Theta Z\approx2\Theta Z,$$ and the perturbed surface
element, $${\delta d{\bf S}\over dS}= \varepsilon
\left(2\Theta,-\frac{\partial\Theta}{\partial\theta},-\frac{{\rm
i}m\Theta}{\sin\theta}\right) Z,$$ where $dS=R^2d\mu d\varphi$.
The tangential components arise from the change of the normal to
the stellar surface. These components are needed also in the
expression for the perturbed limb-darkening, $\delta h_x$. The
perturbed flux is thus given by
\begin{equation}
\delta L_x=\int_S[({\delta\cal F}_x h_x
+{\cal F}_x \delta h_x)dS\ve_r+{\cal F}_x h_x\delta d{\bf S}]\cdot\vn_{\rm obs},
\end{equation}
where $x$ represents a photometric band, $\vn_{\rm obs}$
is the unit vector toward observer, and the integration is done over
the visible unperturbed hemisphere.
The limb-darkening coefficient is normalized, so that
$$ L_x=\int_S{\cal F}_x h_x\tilde\mu dS,$$
where $\tilde\mu=\vn_{\rm obs}\cdot\ve_r$.
Static atmosphere models are used to
calculate wavelength-dependent flux, ${\cal F}_x(T_{\rm eff},g)$,
and the limb-darkening coefficient, $h_x(T_{\rm eff},g)$.

The disc-averaged change of radial velocity is given by
the expression
\begin{equation}
<\!v_{\rm rad}\!>= L_x^{-1}\int_S (\vv_{\rm puls}+\vv_{\rm rot})\cdot{\bf n}_{\rm obs}
{\cal F}_xh_x~ {\bf n}_{\rm obs}\cdot d{\bf S},
\end{equation}
where
$$\vv_{\rm puls}={d\vxi\over dt}\quad
\mbox{ and }\quad \vv_{\rm rot}=R\Omega\sin\vartheta\ve_\varphi$$
In the linear approximation, the contribution from the first term
may be calculated with unperturbed $h$, ${\cal F}$, and ${\bf S}$.
The contribution from the second term arises
solely from perturbation of these quantities. Thus, the radial
velocity change is given by

\begin{eqnarray}
<\!v_{\rm rad}\!>&=&\int_S\biggl\{\left[\vv_{\rm puls}+
\left({\delta{\cal F}_x\over{\cal F}_x}+{\delta h_x\over h_x}\right)
\vv_{\rm rot}\right]\cdot{\bf n}_{\rm obs}\tilde\mu+
\nonumber\\&&
 (\vv_{\rm rot}\cdot{\bf n}_{\rm obs}){\delta d{\bf S}\over dS}\cdot{\bf n}_{\rm obs}\biggr\}{\cal F}_xh_x dS.
\end{eqnarray}

We show in Fig.\,2 the amplitudes of light and radial velocity as functions of the inclination angle for the
modes listed in Table 1. These amplitudes are calculated with the arbitrary choice $|\varepsilon|=0.01$ and
for the Geneva $V$ band, which is similar to Johnson's $V$. The plotted quantities are
$$A_V=1.086\left|\delta L_x\over L_x\right|\qquad\mbox{ and}\qquad
A_{\rm vrad}=|<\!v_{\rm rad}\!>|.$$

In the dependence $A_V(i)$, we see certain patterns which are the
same as for modes described by single spherical harmonics. Zonal
modes are best seen from near polar direction, while symmetric
(even) modes are best seen from near equatorial direction. There
are, however, significant differences. In particular, an increase
of $\lambda$ does not lead to so large amplitude reduction as
expected from the $\ell(\ell+1)\rightarrow\lambda$ replacement.
Compare, for instance, two sectorial $\ell=1$ modes. There is the
factor 25 difference in $\lambda$ between $m=-1$ and $m=+1$ modes
and only the factor 4 in $A_V$, while the factor $\sim12$ would be expected
in the non-rotating case, on the basis of the decline of the disc-averaging
factor, $b_\ell$, with $\ell$ (e.g. Dziembowski 1977).
According to our calculations the $m=+1$ mode is stable.
Among unstable modes, the $(\ell=1, m=-1)$ is the second best
visible, after $(\ell=2, m=+2)$, from the near equatorial
directions. Despite of near-equatorial trapping caused by
rotation, the optimum inclination for detection of tesseral modes,
such as $(\ell=2, m=\pm1)$, is shifted toward the pole because
then cancellation of contributions from the two hemispheres, which
have opposite signs, is reduced. Such a cancellation is
responsible for rather low amplitudes of the r modes.

The behaviour of the radial velocity amplitudes, depicted in the lower
panels of Fig.\,2, shows rather different pattern. At
intermediate inclinations of the rotation axis, the r modes turn
out most easily detectable. At large inclinations (near equatorial
observer) the modes $|m|=\ell$ are best seen both in light and
radial velocity variations. However, the $ $ ratio exhibits a
strong mode dependence. Interpretation of the radial velocity
amplitude pattern is complicated because the contribution from
pulsation and rotation are of comparable size. Here we would like
only to stress that the strong mode dependence of the amplitude
ratios points to the good prospect for mode identification by
combining spectroscopy and photometry data.
\begin{figure}
\centering
\includegraphics[width=88mm, clip]{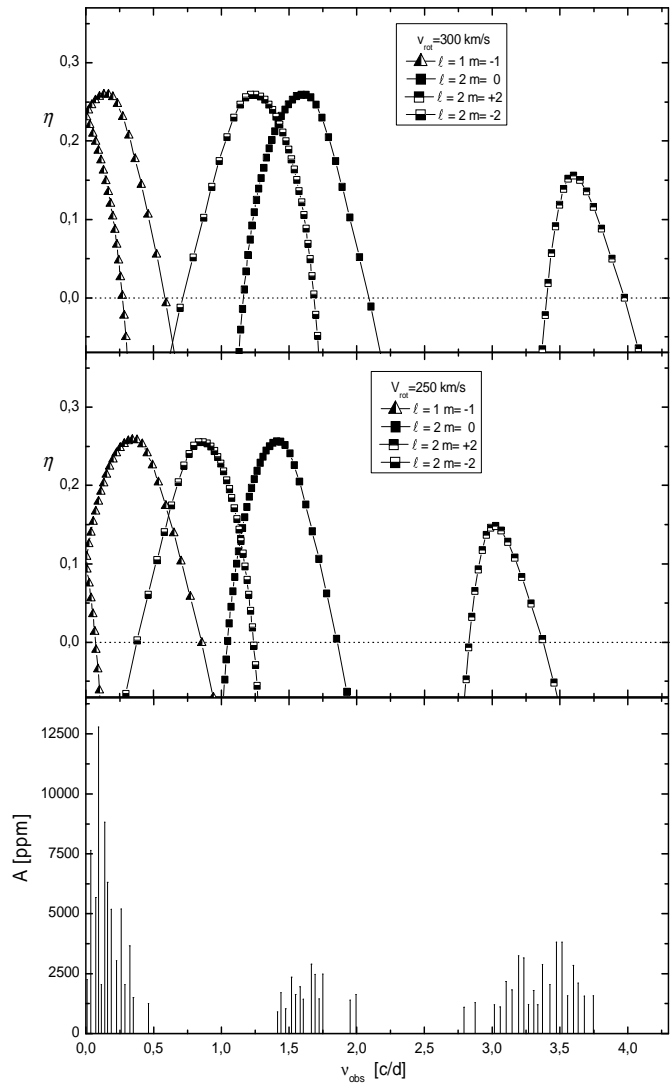}
\caption{ Growth rates for symmetric (even) modes, which are preferentially visible
from the near equatorial directions (see Fig. 2), in models of HD
163868 calculated with two indicated equatorial velocities of
rotation (the upper panels) and amplitudes the peaks detected in
the {\it MOST} data.}
\end{figure}

\section{Peaks in the HD 163868 oscillation spectrum}

The measured value of $v\sin i$ in HD 163868 is high. Therefore, we assume
that the star is seen from a near equatorial direction and, as
possible candidates for identification, we consider only symmetric modes.

There are three groups of frequencies found by Walker et
al.\,(2005) in the star. That with $\nu_{\rm obs}<0.5$ c/d could
only be interpreted in terms of the retrograde $\ell=1$ modes.
Possible interpretations for the group with $\nu_{\rm obs}$
between 1.4 and 2 c/d are $\ell=2$ modes with $m=-2$ and $m=0$.
For the group with $\nu_{\rm obs}$ between 2.8 and 3.75 the only
interpretation is $\ell=2, m=+2$. In this last case our mode
identification is the same as that of Walker et al.\,(2005).

In Fig.\,3 we see that our interpretation is also not fully
satisfactory though, unlike our predecessors, we have no
difficulty with interpretation of lowest frequency group, where
the highest amplitudes are found. Just like theirs, our predicted
range at the highest frequencies is somewhat narrow. The largest
discrepancy between results of our calculation and observation is
for the intermediate group. Certainly, the discrepancy may be
reduced by adjusting  $v_{\rm rot}$ and $T_{\rm eff}$.

It is possible that the some of the peaks in the HD 163868 osillation
spectrum, in particular those in
the low frequency range, are artefacts of the data processing.
Interpretation of peaks in terms of specific modes should be
tested with simultaneous multiband photometry and radial
velocity data. We have seen in the previous section that the ratio of
radial velocity to light amplitude is strongly mode dependent and
it is exceptionally large for the retrograde $\ell=1$ modes, which
may explain peaks in the low frequency range. The dominant peak in
this range and in the whole oscillation spectrum has photometric
amplitude of 0.013 mag in the {\it MOST} photometric system.
Assuming that the amplitude in the $V$ band is similar and using plots in
Fig.\,2, we find radial velocity amplitude of about 30 km/s, which
should be easily measured in radial velocity data spanning
sufficiently long time interval.

\section{Conclusions}

Data on oscillations in Be stars may be helpful
in explaining activity of these stars. The application is twofold.
Firstly, the data may be a source of seismic information on
internal rotation. Secondly, the data may allow to assess the role
of the feedback effect of oscillation in these stars.
Both aims require identification of modes detected in the observed
oscillation spectra. The clue to identification which may be provided
by the theory is the determination of a set of modes that
may be excited and detected.

Effects of rotation, which are important in Be star oscillations,
must be taken into account. This is most easily done with the
traditional approximation. Validity of this
approximation in application to slow modes in Be stars, we
discussed in Section 2. We applied our non-adiabatic version of
this approximation to a model of HD 163868, the first Be star with
the rich oscillation spectrum. The spectrum, determined from the
{\it MOST} photometric satellite data, and its preliminary partial
interpretation, has been recently published by Walker et
al.\,(2005).

We showed that there is a large number of unstable modes in the
model of this star, covering the whole frequency range of the peaks
in the HD 163868 spectrum and beyond. The instability extends to
modes of high horizontal degrees, corresponding to $\ell\approx24$ in the
case of no rotation. We found no significant difference in
driving between prograde and retrograde modes.

Our calculations of the relative visibility of the unstable modes
show that the three consecutive groups of the peaks may be
explained by low-degree high-order g modes seen from the near
equatorial direction. Specifically, the group with frequencies
below 0.5\,c/d  may be explained by retrograde $\ell=1$ modes, the
group around 1.7\,c/d by retrograde sectorial and zonal $\ell=2$
modes the group around 3.5\,c/d by prograde sectorial $\ell=2$
modes. Only the interpretation of the highest frequency group is
the same as that proposed by Walker et al. They interpret the
intermediate group in terms of $\ell=1$ prograde mode, which we
find stable. In contrast, the modes that we associate with this
group are found stable in their calculations. Also they found all
retrograde $\ell=1$ modes to be stable and therefore did not find
a satisfactory interpretation for the lowest frequency group,
where peaks have the highest amplitudes. They proposed that some
of the peaks in this group may be interpreted in terms of
retrograde r modes, which are found unstable, both by them and us,
but their interpretation leaves unexplained highest peaks and
there is also a problem with the visibility of such modes from the
nearly equatorial direction. A possible cause of the difference in
interpretation of the same data is the use of very different
approximations in the treatment of the difficult problem of slow
oscillations in rotating stars.

\section*{Acknowledgments}
We thank the referee, Rich Townsend, for his comments which have
significantly helped us to improve the manuscript, and Hideyuki Saio for
a conversation that also had an impact on the final version of
this paper.
This investigation has been supported
by the Polish MNiI grant No.~1~P03D~021~28.

\end{document}